\documentstyle[aps,pre,psfig]{revtex}

\begin{document}
\twocolumn

\title{Motion, Creation and Annihilation of Disclinations\\
       in Multi-Domain Structured Nematic Liquid Crystal Cells}

\author{M.~Reichenstein$^{*}$,  H.~Stark$^{\text{\ddag}}$,
        J.~Stelzer$^{\text{\dag}}$, and H.-R.~Trebin$^{*}$}
\address{$^{*}$Institut f\"ur Theoretische und Angewandte Physik,
         Universit\"at Stuttgart,\\ Pfaffenwaldring 57,
         D-70550 Stuttgart, Germany \\
         $^{\text{\ddag}}$Fachbereich Physik, Universit\"at Konstanz,
         78457 Konstanz, Germany\\
         $^{\text{\dag}}$Schwetzinger Str.~20, 69190 Walldorf (Baden),
         Germany}

\maketitle

\begin{abstract}
We study dynamical processes in a multi-domain (MD) structured
nematic liquid crystal cell with a particular emphasis on the motion,
creation and annihilation of disclinations.
In the MD cell right- and left-handed director helices alternate due
to a special choice of the director pretilt angles at the surfaces. As
a result, a net of twist disclinations occurs. We have implemented a
numerical algorithm based on a pure rotational dynamics of the
director field to monitor the motion of the defect lines during the
switching process, {\em i.e.}, when an electric voltage is applied to
or removed from the cell. We demonstrate that the total light transmission
versus time is not affected by the presence of the defects compared to a
conventional twisted nematic (TN) cell.
If the pretilt angles at the surfaces
are sufficiently small, the twisting sense of one species of helices is
reversed and a configuration free of defects occurs. On the other
hand, for an applied voltage twist disclinations close to the surface
have to exist. Therefore, defect lines are created or they annihilate
during the switching process. We investigate these situations
in detail and reveal the underlying mechanisms.
\end{abstract}


\section{INTRODUCTION} \label{sec:Intro}

Liquid crystal displays (LCDs) take a very important position
in modern information technology (IT). They are economical both in
their
consumption of energy and their request of space. Recent developments
including
optical compensation and in-plane switching
(IPS)\ \cite{itapdb:OhE1995} technology
improve the quality of high-end LCDs applied in monitors. While these
techniques are too expensive for the production of cheap, low-end displays,
simpler solutions for one of the main problems are required: the contrast
of conventional displays based on the twisted nematic (TN)
cell\ \cite{itapdb:Schadt1971} strongly depends on the viewing angle. The
multidomain cell
(see\ \cite{itapdb:Schadt1996,itapdb:Chen1996} and references cited in
these articles) is a very simple approach to reduce this deficiency.

In the classical TN cell the liquid crystal molecules at the two confining
glass plates are anchored parallel to the surface. They enclose an angle
of $90^{\circ}$ so that the molecules in the bulk form a helix. By
slightly tilting the molecules at the surface, the twisting sense of the
helix can unambiguously be chosen left or right.
In the standard TN cell, the molecules in the bulk are always completely
aligned along the electric field by applying a sufficiently high
voltage. In a TN greyscale display, different grey values are achieved
by varying the applied voltage. The molecules are not fully aligned
along the electric field. Their direction is governed by an
equilibrium between elastic and electric torques.
The helical structure is still partly recognizable.
Since all the helices have the same twisting sense,
the inversion symmetry of the TN cell is broken. As a result
the contrast strongly depends on the viewing angle.

The basic idea of the multi-domain (MD) concept is to avoid this
symmetry breaking. The two surfaces of the cell are treated such that
alternating domains of left- and right-handed helices occur like on
a chessboard\ (see Fig.\ \ref{fig:Schematic}).
The optical effects of the different domains compensate each other at
the length scale of the display pixel, and hence the dependence of the
contrast on the viewing angle is strongly reduced.
Different surface treatments are employed to realize the MD concept\
\cite{itapdb:Chen1995a,itapdb:Schadt1996,itapdb:Chen1996,itapdb:Li1996b}.
The improvement in the contrast is clearly visible.

Since helices of different twisting sense do not match,
all realizations of the MD cell exhibit a characteristic net of line defects
situated at the border of the domains.
The radial extension of the disclination core is of the order of
10$\,$nm, {\em i.e.}, much smaller than the width of one domain
which amounts to 100$\,\mu$m.
Therefore, the disclinations can be neglected for the contrast
properties of the display.

However the occurrence of the disclination lines depends on the choice of
the
tilt angle of the molecules at the surface. The free energy of a
disclination line consists of a core and an elastic part. By reversing
the twisting sense of one species of helices, the disclinations are
eliminated.

\begin{figure}
\centerline{\psfig{figure=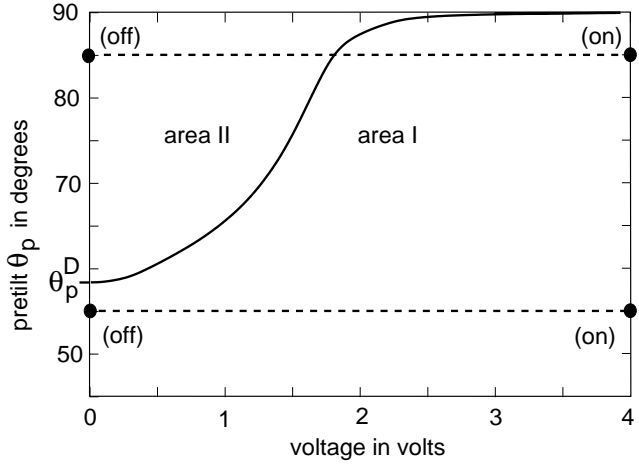,width=8.5cm}}
\vspace*{0.3cm}
\caption[]{Configuration phase diagram of the multi-domain cell
  reproduced from Ref.\ \cite{itapdb:Reichenstein1999}.
         Two areas can be distinguished. Area I: Disclination lines
         are present as expected due to the boundary conditions. Area II:
         For low voltages and pretilt angles close to planar anchoring, the
         disclination lines are removed due to the inverse process of
         splay canceling. All
         helices possess a common twisting sense.
         Two switching processes are investigated in the following,
         one along the upper path at $\theta_p=85^{\circ}$ and a
         second one in analogy to the lower path.}
\label{fig:Phasediag}
\end{figure}

 To accommodate the helices with the unfavorable
twisting sense, additional splay deformations close to the surfaces
have to be introduced. If their elastic free energy is less than the core
energy of the disclination lines, the defect free configuration
occurs. The reverse phenomena, the avoidance of splay deformations by
introducing alternative distortions or even disclinations as in the
case of the MD cell, is denoted
{\em splay canceling}\
\cite{itapdb:Press1975,itapdb:Lavrentovich1990c,itapdb:Chen1996}.
It appears for surface tilt angles $\theta_{p}$
smaller than a maximum value $\theta_{p}^{D}$,
as illustrated in the configuration
phase diagram of Fig.~\ref{fig:Phasediag}. Note that we measure the tilt
angle with respect to the normal of the plates. Then
$\theta_{p} = 90^{\circ}$ means that the molcecules are parallel to the
surface. The figure is reproduced from Ref.\ \cite{itapdb:Reichenstein1999}
where we investigated splay cancelling by numerical means.

When the MD technique is used in displays, the surface anchoring
has to be chosen in such a way that the chessboard pattern of left- and
right-handed helices and therefore the disclination lines are always
present, particularly at zero voltage. This corresponds to
the lower path in Fig.~\ref{fig:Phasediag}.
In section\ \ref{subsec:display}
we will discuss the switching process of such a cell.

From a theoretical point of view the upper path in Fig.~\ref{fig:Phasediag}
is of greater interest. The
configuration at $U=0\ \mathrm{V}$ consists of helices with a common
twisting sense whereas the director field in area I contains disclination
lines. Therefore, during the switching process a configurational transition
must take place where disclinations are created or disappear.
In section\ \ref{subsec:switchonnoff}, we examine the switching process along the upper dotted line.
We investigate in detail how disclinations are nucleated via ring
defects and how they are removed when line defects cross each other.

The paper is organized as follows.
In Section~\ref{sec:TheoryNumerics} we introduce the geometry
of our problem, summarize the theory, and introduce the numerical methods
to investigate the dynamical properties of the MD cell.
In addition, we provide an automated search for disclinations and
explain the visualization of the transmitted light.
In Section~\ref{sec:SwitchConfTrans} the results of the simulations are
presented, and we finish with conclusions in Section~\ref{sec:Conclusions}.

\section{Geometry, Theory and  Numerical  Algorithms}
\label{sec:TheoryNumerics}

\subsection{Geometry}

\begin{figure}
\centerline{\psfig{figure=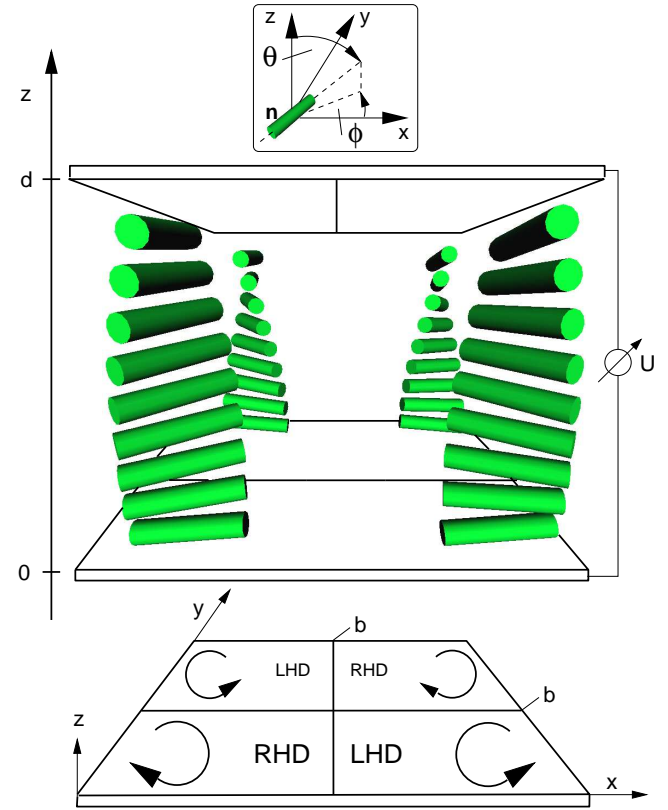,width=8.5cm}}
\vspace*{0.3cm}
\caption{Schematic view of the multi-domain cell configuration studied
  in this article:
         The bottom and top plates are situated at $z=0$ and $z=d$.
         They are divided into stripes of width $b$ along the $x$
         or the $y$ axis, respectively.
         The nematic director is described by spherical coordinates
         $\phi$ (polar or twist angle)
         and $\theta$ (azimuthal or tilt angle). The anchoring at the
         plates is such that the director at the surface is aligned
         along the stripes. Its tilt angle $\theta_{p}$ is fixed, however the
         twist angle $\phi_{p}$ varies by $180^{\circ}$ when going
         from one stripe to the other as summarized in
         Table~\ref{tab:coords}. As a result of such an anchoring structure,
         a chessboard pattern of domains with left-handed ({\em LHD})
         and right-handed ({\em RHD}) helices is expected.}
\label{fig:Schematic}
\end{figure}

In our simulations we use a realization of the multi-domain cell
introduced by
Schadt {\em et al.}\ \cite{itapdb:Schadt1996}. Figure\ \ref{fig:Schematic}
gives a schematic view of the cell. Cartesian coordinates are used to
describe the cell geometry. The average direction of the molecules
is characterized by the director ${\mathbf n}$. Since it is a unit
vector, we employ a twist ($\phi$) and tilt ($\theta$) angle to
parametrize its Cartesian components:
$n_{x} = \text{sin}\theta\, \text{cos} \phi$,
$n_{y} = \text{sin} \theta\, \text{sin} \phi$, and
$n_{z} = \text{cos}\theta$. Due to a special surface treatment the lower
and upper glass plates are divided into stripes pointing along the
$x$ or $y$ axis, respectively.

At both plates the surface tilt
angle $\theta_p$ assumes a common value during a simulation run,
while the surface twist angle $\phi_p$ varies for each stripe according to
Table~\ref{tab:coords}.
The result of such a surface anchoring
structure are domains of director helices with alternating twisting sense.
Since helices of opposite twisting sense do not match, a net of
crossed twist disclinations should occur.

\begin{figure}
\centerline{\psfig{figure=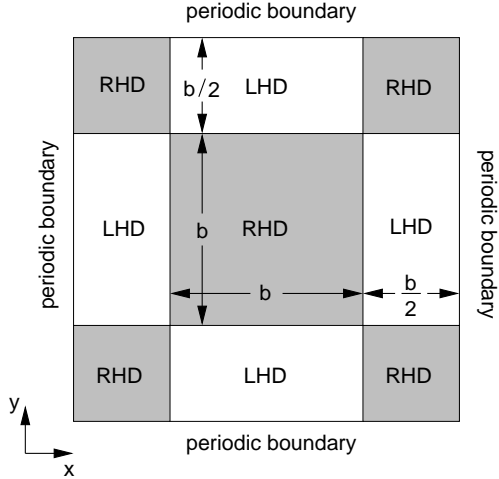,width=6.5cm}}
\vspace*{0.3cm}
\caption[]{Simulation box for the MD cell. Compared to Fig.\
     \ref{fig:Schematic}, the unit cell is shifted by $b/2$ along the
     $x$ and $y$ direction to avoid disclination lines at the boundary
     of the box.}
\label{fig:Schematic_from_top}
\end{figure}

The unit cell of our simulation box has a width and depth of $2 b$, and
its thickness is $d$. We employ periodic boundary conditions along the
$x$ and $y$ direction. To avoid disclinations at the boundary of
the simulation box, we shifted the unit cell by $b/2$ along the
$x$ and $y$ direction compared to Fig.\ \ref{fig:Schematic}.
The final simulation box is illustrated in
Fig.~\ref{fig:Schematic_from_top}.
For the lengths we chose $b = 9 \mu\mathrm{m}$ and $d = 6 \mu\mathrm{m}$.

\subsection{Equations and their Numerical Solution}
\label{subsec:Equations}

To obtain an initial configuration for our dynamic simulations, we
first minimize the total free energy $F$ of the simulation box. It
consists of two parts: the elastic or Frank free energy and
the electric free energy:

\begin{equation}
F= \int[ f_{\text{elast}}({\mathbf x})
      + f_{\text{elec}}({\mathbf x})]\ \text{d}^3 x \enspace.
\end{equation}
The Frank free energy density $f_{\text{elast}}({\mathbf x})$
describes elastic distortions in the director field
${\mathbf n}({\mathbf x})$. It can be divided into three basic modes
with their corresponding elastic constants splay ($k_{11}$), twist
($k_{22}$) and bend ($k_{33}$):
\begin{eqnarray}
\label{equ:freenerN}
f_{\text{elast}}({\mathbf x})
     &=&\frac{1}{2}k_{11}(\text{div}\ {\mathbf n})^{2}
      + \frac{1}{2}k_{22}({\mathbf n}\cdot\text{curl}\ {\mathbf n})^{2}
                                                            \nonumber \\
    &&+ \frac{1}{2}k_{33}({\mathbf n}\times\text{curl}\ {\mathbf
     n})^{2} \enspace.
\end{eqnarray}

In the nematic phase ${\mathbf n}$ and $-{\mathbf n}$ describe the
same situation. This symmetry enables, {\em e.g.}, the existence
of disclination lines. In a numerical simulation, one has to discretize
the director derivatives. In order to avoid artificial
derivatives which occur when neighboring directors are nearly anti-parallel to
each other, they always have to be aligned along each other.
To surpass this difficulty, we rewrite the free energy with the help
of the symmetric and traceless order parameter field
${\mathbf Q}({\mathbf x})$ of second rank. In the uniaxial nematic phase its
Cartesian components are
\begin{equation}
\label{equ:tensorord}
Q_{ij} = n_i n_j -\frac{1}{3}\delta_{ij}, \qquad i,j=1,2,3,
\end{equation}
where $n_i$ is the $i$-th component of the director and $\delta_{ij}$ is the
Kronecker symbol. We do not introduce the Maier-Saupe order parameter $S$
since it is constant in the nematic phase.
With the help of scalar invariants of $Q_{ij}$ and
$Q_{ij,k}$ ($,k$ denotes a spatial derivative along the $x_k$ axis),
the Frank free density can be written in terms of  $Q_{ij}$ and $Q_{ij,k}$
\cite{itapdb:Longa1987}:
\begin{eqnarray}
\label{equ:freenerg}\nonumber
    f_{\text{elast}}({\mathbf x})&=  & \frac{1}{12}(3{k_{22}}-
    k_{11}+k_{33})Q_{ij,k}Q_{ij,k}
    \nonumber\\&+&\frac{1}{2}k_{11}
    Q_{jk,k}Q_{jl,l}
    -\frac{1}{2}k_{22}    Q_{jk,l}Q_{jl,k}
    \nonumber\\&+&\frac{1}{4}(k_{33}-k_{11})
    Q_{jk}Q_{lm,j}Q_{lm,k}
\end{eqnarray}
Substituting Eq.~(\ref{equ:tensorord}) into Eq.~(\ref{equ:freenerg}), the
Frank free energy density (\ref{equ:freenerN}) is recovered.

The voltage applied to the plates of the MD cell induces an electric
field. Its electric free energy in terms of the electric potential $U$
is
\begin{eqnarray}
    f_{\text{elec}}({\mathbf x})=\frac{1}{2}\epsilon_{0}\Delta\epsilon
    \left( \frac{\epsilon_{\|}+2\epsilon_{\bot}}
      {3(\epsilon_{\|}-\epsilon_{\bot})}
      \delta_{ij} + Q_{ij}\right)  U_{,i}U_{,j},
\end{eqnarray}
where $\epsilon_{\|}$ and $\epsilon_{\bot}$ are the eigenvalues of the
dielectric tensor $\epsilon_{ij}=\epsilon_{\bot}\delta_{ij}
+(\epsilon_{\|}-\epsilon_{\bot}) n_{i} n_{j}$. They stand for the
dielectric constants parallel and perpendicular to the director.
The dielectric anisotropy is $\Delta\epsilon=\epsilon_{\|}-\epsilon_{\bot}$.

The minimum of the free energy $F$ is found by solving the
Euler-Lagrange equations for the three independent scalar fields,
{\em i.e.}, the tilt angle $\theta({\mathbf x})$,
the twist angle $\phi({\mathbf x})$, and the electric potential
$U({\mathbf x})$.
With the help of the following chain rule for functional derivatives\
\cite{itapdb:Stark1999}, we can employ the Frank
free energy density $f_{\text{elast}}$ in the ${\mathbf Q}$
representation (see Eq.\ \ref{equ:freenerg}) to set up the
Euler-Lagrange equations:
\begin{eqnarray}
\label{equ:eulerlagrange_PHITETA}
\frac{\delta F}{\delta w}
        &=& \frac{\delta F}{\delta Q_{ij}}\frac{\delta Q_{ij}}{\delta w}
        \nonumber  \\
        &=&\left[\frac{\partial f}{\partial Q_{ij}}
 - \frac{\partial}{\partial x_{m}}
                            \frac{\partial f}{\partial (Q_{ij,m})}\right]
                     \frac{\delta Q_{ij}}{\delta w}=0
\end{eqnarray}
where $w$ stands for $\phi$ or $\theta$. Einstein's summation convention
is used. The Euler-Lagrange equation for $U$ is
\begin{equation}
\label{equ:eulerlagrange_VOLT}
\frac{\delta F}{\delta U} =  \frac{\partial f}{\partial U}
 - \frac{\partial}{\partial x_{m}}\frac{\partial f}{\partial (U_{,m})}
                     =0.
\end{equation}

The Euler-Lagrange equations were discretized on a Cartesian lattice
and a standard Newton-Gau\ss-Seidel procedure was employed
\cite{itapdb:Press1992a} to solve them.
The calculated minimum energy configurations then served as
initial configurations for the dynamic simulations.

To handle the dynamics of a MD cell, we employed a pure rotational
dynamics of the director field without any coupling to the velocity
field \cite{itapdb:Vertogen1988b}:
\begin{equation}
\label{equ:dynamicphiandtheta}
\frac{\delta F}{\delta \theta} =
                - \gamma \frac{\partial \theta}{\partial t}, \qquad
\frac{\delta F}{\delta \phi}   =
             - \gamma \frac{\partial \phi}{\partial t}\sin^{2}\theta,
\end{equation}
where $\gamma$ is a typical rotational viscosity.
They were solved by an explicit time integration method\
\cite{itapdb:Press1992a}. The electric potential $U$
was assumed to adjust instantaneously to the director rotation
time scale, so that the static equation (\ref{equ:eulerlagrange_VOLT})
was solved at each time step.

Throughout all the simulations the  values for the material
parameters were chosen as in \cite{itapdb:Schadt1992}:
$k_{11} = 12.4 \cdot 10^{-12}\, \mathrm{N}$, $k_{22} =  6.5
\cdot 10^{-12}\, \mathrm{N}$, $k_{33} = 19.9 \cdot 10^{-12}\,
\mathrm{N}$, $\epsilon_{\parallel} = 8.03$, $\epsilon_{\perp}$ =
$3.59$, and $\gamma_1 = 17 \cdot 10^{-3}\, \mathrm{Nsm^{-2}}$.

At last we add a note about the disclinations. In our simulation we
did not introduce an extra core energy for reasons of simplicity.
As the length of the disclinations varies, it would have
been too complicated numerically to correct for the core energy at each time
step.

\subsection{Automated Search for Defects and Visualization}
\label{subsec:DefectfinderVisualization}

We developed some tools to visualize the director field configuration
in the MD cell. Especially, we have implemented an algorithm of
automated Burgers circuits to detect disclinations automatically.
The algorithm adopts the conventional definition of a line defect to
a grid of lattice points. To detect a line defect at a certain lattice
point, we move around it on a loop of nearest neighbours.
When moving from one lattice point to the next, the directors are
aligned such that they enclose an angle smaller than $90\,^{\circ}$.
At last, we compare
the directors at the starting and end point of the loop. If they
enclose an angle larger than $90\,^{\circ}$, a line defect is present.
Whenever the angle between two neighboring directors on a loop is
very close to $90^{\circ}$, the search is performed on a loop of
next-nearest neighbors.
To obtain a three-dimensional picture of the position of the disclination
lines, we use the Virtual Reality Modeling Language \cite{itapdb:www2000}.
Each defect position on the grid is marked by a small sphere.
All spheres together present an image of the disclination line.

We also calculated the optical transmission for light incident parallel to
the surface normal of the plates
({\em i.e.}, along the negative $z$ axis) based on the
$2\times 2$ Jones-matrix method\ \cite{itapdb:Drzaic1995}. The incident
light is polarized along the $x$ axis at $z=d$, and it is analysed by a
polarizer along the $y$ axis at $z=0$ after being modified by the
liquid crystal.
The resulting intensity is mapped onto a greyscale. Maximum
transmission corresponds to white and minimum transmission to
black. The resulting grey values for each grid point in the $xy$ plane
represent the transmission pattern of the liquid crystal
configuration. In our calculations we always used monochromatic light of
wavelength $\lambda = 544\, \mathrm{nm}$. The refractive indices were
$n_{\perp} = 1.489$ and $n_{\parallel} = 1.762$.

\section{Switching the MD Cell}
\label{sec:SwitchConfTrans}

In this section we investigate the dynamic processes in the MD cell
which take place during a switching process, {\em i.e}, after an
electric voltage of $4\, \mathrm{V}$ has been applied to or removed
from the cell. We first comment on the configuration as it is used in
displays, where the net of disclinations is always present
(subsection~\ref{subsec:display}). Then we consider the upper
switching path in the phase diagram of Fig.~\ref{fig:Phasediag}.
A configurational transition occurs where line defects have to be
created or removed from the cell (subsection~\ref{subsec:switchonnoff}).

All the simulations presented were performed on a grid with $40$ gridpoints
in $x$, $y$ and $z$ direction, respectively. All switching times referred 
to in the following text and the figures are measured in milli seconds.

\subsection{The MD Cell as a Display} \label{subsec:display}

It would be natural to study the switching process along the lower
path in Fig.\ \ref{fig:Phasediag}. However, we note that the tilt
angle $\theta_{p}^{D} = 58^{\circ}$ below which the disclinations are
always present is much smaller compared to realizations of the MD display.
For example, Schadt {\em et al.\/} work at a tilt angle of
$\theta_{p} = 88.7^{\circ}$\ \cite{itapdb:Schadt1996}. Chen {\em et al.\/}
have studied the occurrence of splay canceling in the MD cell in
detail\ \cite{itapdb:Chen1997a},
and they even fabricated a display with $\theta_{p}^{D}$ as small as
$65^{\circ}$.
With a rough estimate, the authors could show that the value of
$\theta_{p}^{D}$ depends on the ratio of cell dimensions.

If the director helix is
reversed, an additional splay distortion is distributed over the whole
subpixel. Its total energy for two subpixels is approximated as
an average Frank free energy density times the volume $2b^{2}d$, {\em i.e.},
\begin{equation}
F_{S} \simeq K \, \left(\frac{2(\pi / 2 - \theta_{p})}{d}\right)^{2}
\, b^{2}d \enspace.
\end{equation}

The twist disclinations occur if their core energy $F_{d}$
is smaller than $F_{S}$. The core energy is given by
\begin{equation}
F_{d} = 8 K b \enspace,
\end{equation}
where the average Frank constant $K$ approximates the line energy of
a defect core.
The transition between both configurations takes place at
$F_{S} \approx F_{d}$ or
\begin{equation}
\label{estimate}
\theta_{p}^{D} \approx \pi / 2 - \sqrt{2d/b} \enspace.
\end{equation}
With the parameters of our numerical calculations, $d/b = 1/3$,
we find $\theta_{p}^{D} = 43^{\circ}$ which deviates from the
simulated value, $\theta_{p}^{D} = 58^{\circ}$, due to the rough
estimates for the free energies involved.
However, Eq.\ (\ref{estimate}) demonstrates that the
relatively small angle of $58^{\circ}$ results from the
small lateral extension $b = 18 \mu\text{m}$ compared to typical
pixel sizes of $100\,\mu\text{m}$.

\begin{figure}
\centerline{\psfig{figure=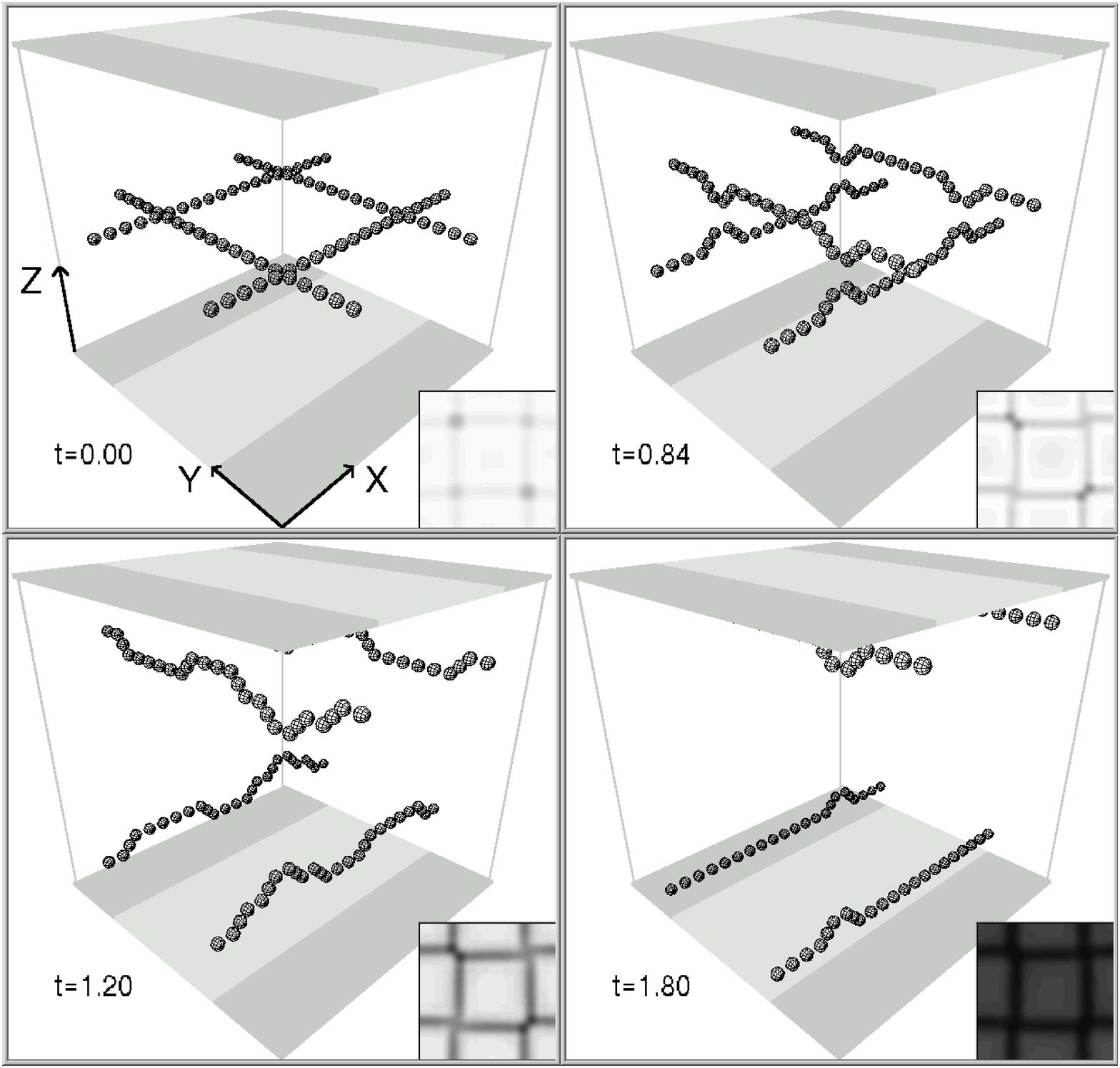,width=8.7cm}}
\vspace*{0.3cm}
\caption{Snapshots of the moving twist disclinations in the MD display
         and its
         transmission of light (see lower right corner) at various
         time steps after switching
         the electric voltage from $U=0\ \mathrm{V}$ to $4\ \mathrm{V}$.
         The upper and lower glass plates are divided into anchoring stripes
         shown in dark and light grey. The director field between the
         two plates is hidden to enlarge the visibility of the
         disclination lines.
         For a detailed description refer to the text.}
\label{fig:array_md_ein}
\end{figure}

To be closer to the experimental realization of the MD display, we
decided to study
the switching process for a cell with pretilt angle
$\theta_{p} = 85^{\circ}$. At zero voltage ($U=0\,\text{V}$)
the defect configuration can nevertheless be prepared since it
corresponds to a metastable state. We construct it by solving the
static equation\ (\ref{equ:eulerlagrange_PHITETA}). Then we set the
voltage to $4\, \mathrm{V}$ and start the time integration algorithm.
For characteristic times in the evolution of the director field,
we plot in Fig.~\ref{fig:array_md_ein} the disclination lines in
the cell and in the lower right corner its light transmission.

\begin{figure}
\centerline{\psfig{figure=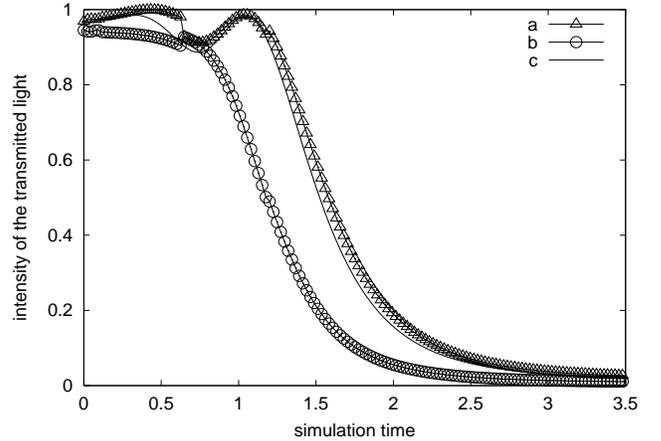,width=8.7cm}}
\vspace*{0.3cm}
\caption[]{Light transmission versus time in the MD display: a) in the
           center of the cell, b) close to a twist disclination, and, for
           comparison, in the TN cell (curve c).}
\label{fig:extract_trans_alongx}
\end{figure}

At $t=0$, the defect lines are situated in the center of the cell
forming a net of twist disclinations. When the voltage is turned on,
the disclinations start to move towards the glass plates (see
Fig.\ \ref{fig:array_md_ein}, $t=0.84$ and $1.20$). At their
intersections they are dipped towards each other, due to an attractive
interaction. When they have reached the bounding surfaces,
they straighten out (see Fig.\ \ref{fig:array_md_ein}, $t=1.80$).
It is instructive to consider the light transmission. At $t=0$, the
disclinations can hardly be seen in the transmission pattern. When
they move towards the glass plates, the contrast becomes
stronger, {\em i.e.}, close to the disclinations the cell switches
faster compared to the rest of the display. This surprising result is
also illustrated in Fig.\ \ref{fig:extract_trans_alongx} where we
plot the transmission versus time close to the defect lines (curve b)
and in the center of the display (curve a). The dip in curve a between
$t=0.5$ and $1$ is a well known phenomenon in a TN cell with
pretilt\ \cite{itapdb:Stelzer1993}.
The transmission characteristic of a TN cell (see curve c) is
also plotted in Fig.\ \ref{fig:extract_trans_alongx}. It
demonstrates that far away from the disclination lines the helices in
the MD cell switch nearly as fast as in a pure TN display. Since the
lateral extension of a pixel is much larger than the core of a
twist disclination, we can conclude that the motion of the defect
lines does not influence the light transmission of a MD display.

We finish with two remarks. First, when the voltage is turned off, the
motion of the disclination lines, as illustrated in the snapshots of
Fig.\ \ref{fig:array_md_ein}, is simply reversed.

Secondly, the position of the twist disclinations as a function of the
pretilt
angle $\theta_{p}$
for zero voltage is the following. Until $\theta_{p} = 80^{\circ}$ the
defect lines are straight and located in the center of the cell. Below
$\theta_{p} = 80^{\circ}$ they start to bend moving towards the glass plates
but still touching each other at the intersections. At
$\theta_{p} = 60^{\circ}$, the lines lose contact, and for further
decreasing $\theta_{p}$ they move closer to the surface.

\subsection{Configurational Transitions in the MD Cell}
\label{subsec:switchonnoff}

\subsubsection{Switching from $0\, \mathrm{V}$ to $4\, \mathrm{V}$}
\label{subsec:switchon}

\begin{figure}
\centerline{\psfig{figure=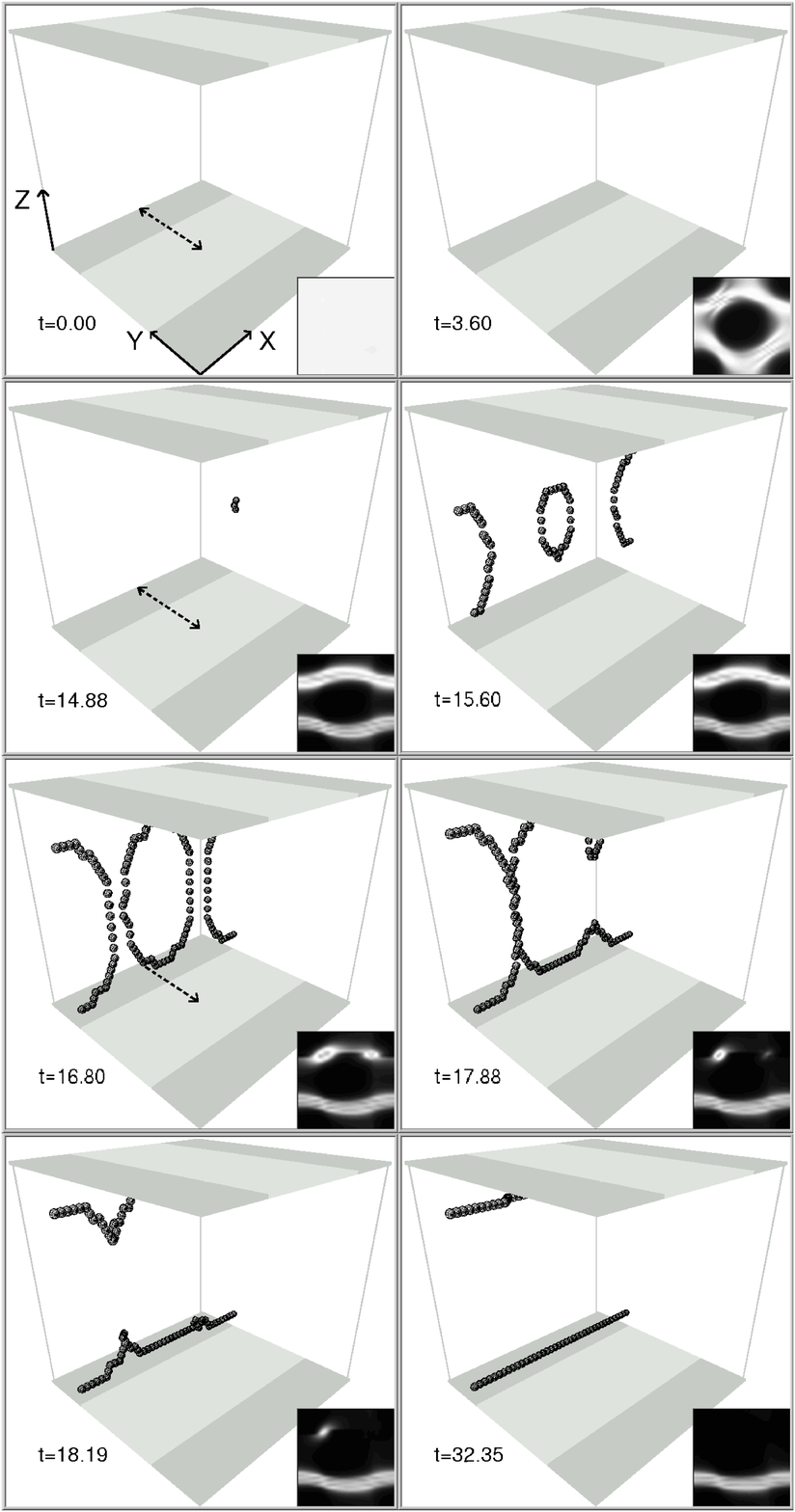,width=8.7cm}}
\vspace*{0.3cm}
\caption{Snapshots of the creation of disclinations in the MD cell and its
         transmission of light (see lower right corner) at various time steps
         after switching the electric voltage from $U=0\ \mathrm{V}$ to
         $4\ \mathrm{V}$.
         The upper and lower glass plates are divided into anchoring stripes
         shown in dark and light grey.
         For a detailed description refer to the text.}
\label{fig:3dsnaps_on}
\end{figure}

For the upper path in Fig.\ \ref{fig:Phasediag}, we calculate the
initial configuration by solving the
static equation\ (\ref{equ:eulerlagrange_PHITETA}) for $0\, \mathrm{V}$.
As before, the voltage is set to $4\, \mathrm{V}$, and the time
integration algorithm is started. Snapshots of the appearing
disclination lines and the light transmission patterns are illustrated
in Fig.~\ref{fig:3dsnaps_on}.

\begin{figure}
\centerline{\psfig{figure=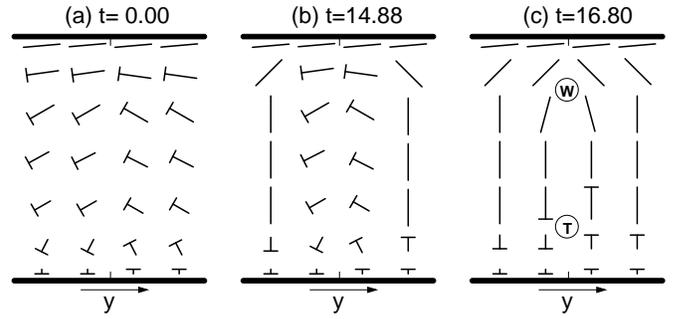,width=8.7cm}}
\vspace*{0.3cm}
\caption[]{Nail picture to illustrate the mechanism for the creation
           of disclination rings in the MD cell. The length of the nail is
           given by the projection of the director on the drawing plane. Its
           head symbolizes the part of the director below the drawing plane.
          a) Undistorted helices, the third and fourth helix have reversed
           their twisting sense by introducing an additional splay deformation.
           The helices belong to a path parallel to the $y$ axis indicated in
           Fig.\ \ref{fig:3dsnaps_on} by an arrow.
          b) After applying an electric field, the first and fourth helix are
           distroyed. c) By introducing a twist (T) and a $+1/2$ wedge (W)
           disclination, the directors in the second and third helix can be
           aligned along the electric field.}
\label{fig:helixsketches}
\end{figure}

At $t=0.0$ all helices have a common twisting sense. The helices
of the RHDs (see Fig.~\ref{fig:Schematic_from_top}) turn in accordance
with their boundary conditions. The helices in the LHDs have reversed
their twisting sense by introducing an additional splay deformation
which is distributed over the whole helix.
No disclinations are present.
Now the voltage of $4\, \mathrm{V}$ is applied and the directors start
to align along the electric field.
At $t=3.6$ a clear difference can be seen between the two types of helices:
In the RHDs the transmission is already close to zero, {\em i.e.}, the
helix is destroyed since the directors are parallel to the electric field.
In the LHDs the transmission is almost unchanged. The helix structure
is still present. We understand that feature from
Fig.\ \ref{fig:helixsketches}(a)
where we present a schematic nail picture of the helices along a path
indicated in the snap shots of the cell in Fig.\ \ref{fig:3dsnaps_on}.
Note that the length of the nail is given by the projection of the
director on the drawing plane and that the head symbolizes the part
of the director below the drawing plane.

In the first and second helix all directors are commonly tilted towards
the electric field, and they can align along the field immediately.
However, the third and fourth helix, whose twisting sense is reversed
by an additional splay deformation, are temporarily stabilized by the
occurrence of a horizontal director.
At $t=14.88$, also the directors in the LHDs are aligned
along the electric field [see fourth helix in Fig.\
\ref{fig:helixsketches}(b)] leaving
two bright stripes in the transmission pattern. They result from
regions symbolized by the second and third helix in
Fig.\ \ref{fig:helixsketches}(b)
whose directors can only be aligned along the electric field by
introducing a $+\frac{1}{2}$ wedge (W) and a twist (T) disclination, as
illustrated in Fig.\ \ref{fig:helixsketches}(c). These disclinations
are part of one of the two defect rings which start to open up in the
$xz$ plane at $t=14.88$ and then grow continuously.
Upon touching each other at $t=17.66$, they
annihilate leaving a wedge and a twist disclination at the upper and
lower plate, respectively. The line defects straighten out until
$t=32.35$. Note that due to the creation of the two defect lines, one
bright stripe has vanished in the transmission pattern.

\begin{figure}
\centerline{\psfig{figure=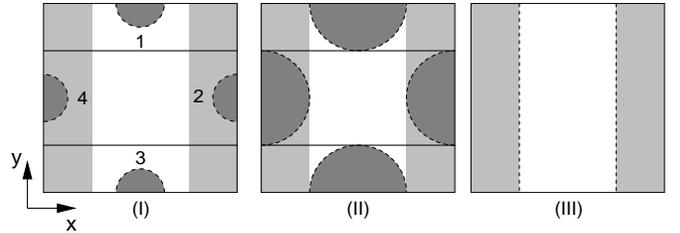,width=8.7cm}}
\vspace*{0.3cm}
\caption[]{Mechanism to rotate the disclination lines close to
the upper plate by $90^{\circ}$. (I) Disclination rings open up in the
regions 1 to 4. (II) They partially annihilate with the existing wedge
disclinations. (III) Defect lines of twist character along the $y$
axis result.  For a detailed description refer to the text.}
\label{fig:finalrings}
\end{figure}

At $t=32.35$ the simulation is trapped in a local minimum. To reach
the global minimum a second pair of disclinations should have been
created to reduce the transmission of the second white stripe. But
even then the cell is not in its gobal minimum configuration since
the disclination lines at the upper plate have to point along the $y$ axis.
To achieve a rotation of the defect lines, we propose the following
mechanism. In Fig.\ \ref{fig:finalrings}(I) the defect lines at the
upper plates
a drawn. Note that in the regions 1 to 4 an additional splay
deformation close to the upper surface is still present. An inspection
of the director field reveals that this splay deformation can be
reduced locally by opening up a disclination rings
[see Fig.\ \ref{fig:finalrings}(I)].
Their right and left parts are twist disclinations whereas the upper
and lower parts have a wedge character. The latter ones annihilate with
the existing wedge disclinations [see Fig.\ \ref{fig:finalrings}(II)],
and the twist parts of the disclination ring produce the two required
defect lines parallel to the $y$ axis.

The final director field of the global minimum consists of two pairs
of disclinations, parallel to the surface anchoring stripes of the
respective glass plates. It is taken as initial configuration in the
following section as illustrated in Fig.~\ref{fig:3dsnaps_off}, $t=0.0$.

\subsubsection{Switching from $4\, \mathrm{V}$ to $0\, \mathrm{V}$}
\label{subsec:switchoff}

\begin{figure}
\centerline{\psfig{figure=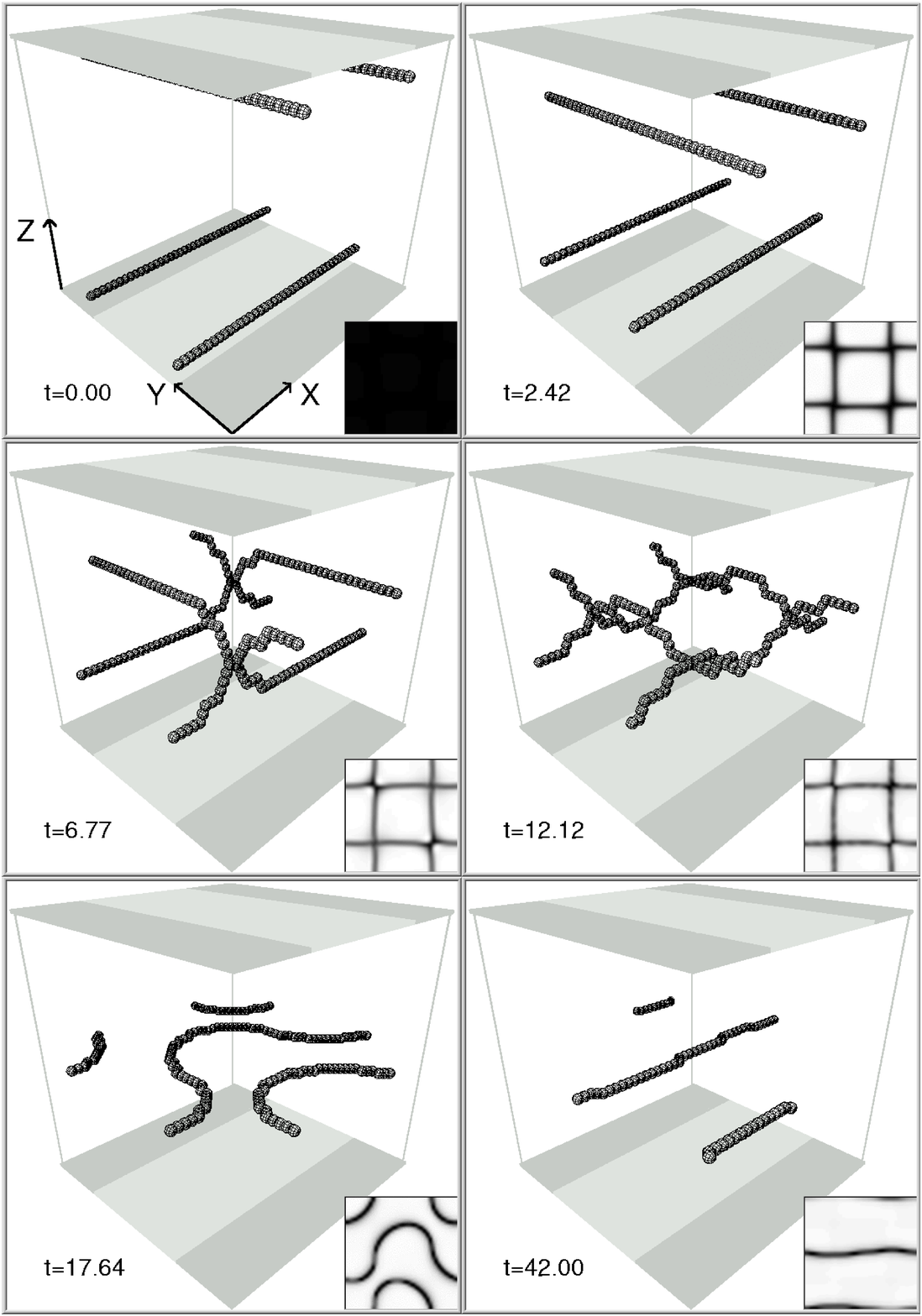,width=8.7cm}}
\vspace*{0.3cm}
\caption[]{Snapshots of the partial disappearance of disclinations in
           the MD cell and its transmission of light (see lower right corner)
           at various time steps after switching the electric voltage from
           $U=4\ \mathrm{V}$ to $0\ \mathrm{V}$.
           The upper and lower glass plates are divided into anchoring stripes
           shown in dark and light grey. For a detailed description
           refer to the text.}
\label{fig:3dsnaps_off}
\end{figure}

In Fig.\ \ref{fig:3dsnaps_off} we illustrate the evolution of the MD
cell after the voltage of $4\, \mathrm{V}$ has been removed at $t=0.0$.
The disclination pattern and the transmission of light are shown.
At $t=0.0$ the directors are still aligned along the $z$ axis parallel
to the electric field direction. The helices are destroyed and the
intensity of the transmitted light is zero.
The twist disclinations are located close to the surfaces.
When the voltage is removed, the director field relaxes due to the
elastic interactions. At $t=2.42$ the helices in the LHDs and RHDs
are restored.

The director configuration has regained its ability to
rotate the polarization of light, and the transmission is almost
at its maximum. Fig.\ \ref{fig:3dsnaps_off}, $t=2.42$, reveals
that the disclinations have moved towards the center of the cell.
They are clearly visible in the transmission.
At $t=5.36$ the disclinations start to dip towards each other due
to a strong attraction in the region of their crossing sections. Then, at
$t=6.77$, recombinations of the disclinations take place which are
completed at $t=12.12$ when a net of disclinations is situated at the
center of the cell at $z=d/2$.

So far, the situation is similar to the
MD display discussed in subsection\ \ref{subsec:display}.
Here, the intersections of the
disclinations eventually break up such that two line defects result
(see Fig.\ \ref{fig:3dsnaps_off}, $t=17.64$) which straighten out
until $t=42.00$. They separate domains with right- and left-handed
helices. Now, our simulation is trapped in a local minimum,
and the configuration does not change any more. The global minimum
would be helices with a common twisting sense. It can be reached
by the following scenario. At $t=12.12$ the intersections of the line
defects should break up such that a ring disclination in the center
of the cell occurs which shrinks and ultimately annihilates leaving
a defect-free configuration. Whether this scenario does not appear in
our simulation due to numerical inaccuracies or whether it is very
unlikely to happen since the intersections have to break up in a
very defined way, has to be answered by an experimental investigation.

\section{Conclusions}
\label{sec:Conclusions}

The concept for the MD cell emerged from the need to improve the
viewing angle of the conventional liquid crystal display based on the
TN cell. In this paper we have studied one realization of the MD cell
in detail. By tilting the director against the planar orientation at
the surface, a chessboard pattern of alternating left- and
right-handed helices is expected where topology requires the existence
of twist disclinations. For sufficiently small tilt angles
$\theta_p$ against the surface normal, the pattern exists. However,
if $\theta_p$ is increased, the twisting sense of one species of
helices is reversed at the cost of an additional splay deformation.
Following Chen {\em et al.} \cite{itapdb:Chen1997a}, we have given
an estimate for the tilt angle $\theta_p^D$ where a transition between both
configurations occurs. It solely depends on the ratio of the cell
dimensions.

We have implemented a numerical algorithm based on a pure rotational
dynamics of the director field to study the dynamics of
the MD cell which is driven by an electric field. Furthermore, we have
developed tools to visualize the defect dynamics and the transmission
of light. In the realization of the
display, where the line defects are always present, we find that the
total transmission versus time is not
affected compared to a conventional TN cell. Although, as a surprising
result,
the switching time close to the disclinations is smaller.
From a theoretical point of view, the second configuration without
defects at zero voltage is more appealing. It provides a playground
for studying the creation and annihilation of line defects. We have
investigated these processes for the MD cell in detail,
and we have revealed their mechanisms.
E.g., when the voltage is turned on, defect rings with a mixed twist
and wedge character are nucleated. They partly annihilate to create
two twist disclinations at the surface. An obstacle of our
simulations is that the dynamics becomes trapped in a metastable
minimum and does not relax into the global minimum. Therefore, to
check our results, an experimental investigation would be helpful.

M.\ R. gratefully acknowledges a grant from the Sonderforschungsbereich
382 of the Deutsche Forschungsgemeinschaft. H.\ S. is generously
supported by the Deutsche Forschungsgemeinschaft through a Heisenberg
scholarship.



\begin{thebibliography}{10}

\bibitem{itapdb:OhE1995}
M.~Oh-e and K.~{Kondo}.
\newblock Appl.\ Phys.\ Lett. {\bf 67} 3895 (1995).

\bibitem{itapdb:Schadt1971}
M.~{Schadt} and W.~{Helfrich}.
\newblock Appl.\ Phys.\ Lett. {\bf 18}, 127 (1971).

\bibitem{itapdb:Schadt1996}
M.~{Schadt}, H.~{Seiberle}, and A.~{Schuster}.
\newblock Nature {\bf 318}, 212 (1996).

\bibitem{itapdb:Chen1996}
J.~{Chen}, P.~J. {Bos}, D.~L. {Johnson}, {\it et~al.\/}.
\newblock Jpn.\ J.\ Appl.\ Phys. {\bf 35}, 558 (1996).

\bibitem{itapdb:Chen1995a}
J.~{Chen}, P.~J. {Bos}, D.~R. {Bryant}, D.~L. {Johnson}, {\it et~al.\/}.
\newblock Appl.\ Phys.\ Lett. {\bf 67}, 1990 (1995).

\bibitem{itapdb:Li1996b}
J.~{Li}, E.~S. {Lee}, H.~{Vithana}, and P.~J. {Bos}.
\newblock Jpn.\ J.\ Appl.\ Phys. {\bf 35}, 1446 (1996).

\bibitem{itapdb:Press1975}
M.~J. {Press} and A.~S. {Arrott}.
\newblock J.~Phys. (Paris)\ Coll. C1 {\bf 36}, 177 (1975).

\bibitem{itapdb:Lavrentovich1990c}
O.~D. {Lavrentovich} and Y.~{Nastishin}.
\newblock Europhys.\ Lett. {\bf 12}, 135 (1990).

\bibitem{itapdb:Reichenstein1999}
M.~{Reichenstein}, T.~{Seitz}, and H.-R. {Trebin}.
\newblock Mol.\ Cryst.\ Liq.\ Cryst. {\bf 330}, 549 (1999).

\bibitem{itapdb:Longa1987}
L.~{Longa}, D.~{Monselesan}, and H.~R. {Trebin}.
\newblock Liq.\ Cryst. {\bf 2}, 769 (1987).

\bibitem{itapdb:Stark1999}
H.~{Stark}.
\newblock Eur.\ Phys.\ J.\ B {\bf 10}, 311 (1999).

\bibitem{itapdb:Press1992a}
W.~H. {Press}.
\newblock {\it Numerical Recipes in C: The Art of Scientific Computing\/}.
\newblock Academic Press, Cambridge, 2nd edition (1992).

\bibitem{itapdb:Vertogen1988b}
G.~{Vertogen} and W.~H. {de Jeu}.
\newblock {\it Thermotropic Liquid Cristals, Fundamentals\/}.
\newblock Springer-Verlag (1988).

\bibitem{itapdb:Schadt1992}
M.~{Schadt}, K.~{Schmitt}, V.~{Kozinkov}, and V.~{Chigrinov}.
\newblock Jpn.\ J.\ Appl.\ Phys., Pt.~1 {\bf 31}, 2155 (1992).

\bibitem{itapdb:www2000}
{web3D Consortium}.
\newblock {\it {http://www.vrml.org}\/}.

\bibitem{itapdb:Drzaic1995}
P.~S. {Drzaic}.
\newblock {\it Liquid {C}rystal {D}ispersions\/}, volume~1 of {\it Series on
  Liquid Crystals\/}.
\newblock World Scientific Publishing, Singapore (1995).

\bibitem{itapdb:Chen1997a}
J.~{Chen}, J.~{Li}, D.~L. {Johnson}, and P.~J. {Bos}.
\newblock Mol.\ Cryst.\ Liq.\ Cryst. {\bf 302}, 151 (1997).

\bibitem{itapdb:Stelzer1993}
J.~{Stelzer}, R.~{Hirning}, and H.-R. {Trebin}.
\newblock J.~Appl.\ Phys. {\bf 74}, 6046 (1993).

\end{thebibliography}


\narrowtext
\begin{table}
\begin{tabular}{cccc}
     coordinates & surface twist & surface tilt & applied     \\
                 &     angle     &    angle     & voltage     \\ \hline
          $z=0$, $0<y<b\ $ & $\phi_{p}=0^{\circ}\ \ $   & $\theta_p$
           & $U=0\,\mathrm{V}$ \\
     $z=0$, $b<y<2b$ & $\phi_{p}=180^{\circ}$ & $\theta_p$ &
     $U=0\,\mathrm{V}$ \\
     $z=d$, $0<x<b\ $ & $\phi_{p}=90^{\circ}\ $  & $\theta_p$ &
        $U=0 \, \text{or} \, 4 \, \mathrm{V}$  \\
     $z=d$, $b<x<2b$ & $\phi_{p}=270^{\circ}$ & $\theta_p$ &
     $U=0 \, \text{or} \, 4 \, \mathrm{V}$
\end{tabular}
\vspace*{0.3cm}
\caption{Surface anchoring structure for the cell in
  Fig.~\ref{fig:Schematic} and the electric potential $U$ at the surfaces.
 The quantities $\theta_p$ and $U$ are parameters in the simulation.}
\label{tab:coords}
\end{table}

\newpage

\newpage

\end{document}